\documentclass[aps,prd,11pt,tightenlines,nofootinbib]{revtex4}
\usepackage{anysize}
\marginsize{1.5cm}{1.5cm}{1.5cm}{1.5cm}

\usepackage{amsmath,amssymb}
\usepackage{slashed}	

\usepackage{graphicx}
\usepackage{fancybox,color}

\usepackage[active]{srcltx} 
\usepackage[utf8]{inputenc} 
\usepackage[T1]{fontenc} 

\usepackage[normalem]{ulem}

\usepackage{enumitem}
\setitemize{noitemsep,topsep=-0.0em,parsep=0em,partopsep=0em}
\setenumerate{noitemsep,topsep=-0.0em,parsep=0em,partopsep=0em}

\usepackage[font=small,labelfont=bf]{caption}

\let\oldbibliography\thebibliography
\renewcommand{\thebibliography}[1]{\oldbibliography{#1}\setlength{\itemsep}{-3pt}}

\usepackage{xspace}

\newif\iffinal
  \finaltrue		
\newif\ifmarginnote
  \marginnotefalse  

\iffinal
  \marginnotefalse
  
  \def\finaloff#1{}
\else
  
  \def\finaloff#1{#1}
\fi

\ifmarginnote
  \newcommand{\mnote}[1]{\marginpar{\fbox{\small\textit{#1}}}}       
  \newcommand{\parcom}[1]{\hspace{-1.5mm}\marginpar{\flushleft{\small\em #1}}} 
\else
  \newcommand{\mnote}[1]{}        
  \newcommand{\parcom}[1]{}        
\fi

\newcommand{\mybox}[2][blue]{{\color{#1}\fbox{\color{black} #2}}}
\newlength{\myboxL}

\definecolor{mygray}{gray}{0.5}
\newcommand{\mycomment}[2][]{}

\newcommand{\xcite}[1][...]{\mybox{cite #1}}

\def\ra{\rightarrow}
\def\tra{$\;\rightarrow\;$ }

\def\b{\indent $\bullet$\ }

\def\b#1{\textbf{#1}}

\def\>{\rangle}
\def\<{\langle}

\def\Z{\mathbb{Z}}

\def\v{\vec}
\def\d{\partial}
\def\dd{\bar{\partial}}
\def\h{\hat}

\def\s{{}^\dagger}
\def\B{\mathcal{B}}

\def\eps{\varepsilon}
\def\la{\lambda}

\def\Lap{\Delta}

\begin{document}


\title{
Curved spacetimes in the lab
}

\author{Nikodem Szpak\\[0.5em]{\small e-mail: \texttt{nikodem.szpak@uni-due.de}}}

\address{Faculty of Physics, University Duisburg-Essen, Lotharstr. 1, 47057 Duisburg, Germany}



\date{March 31, 2014}

\begin{abstract}
  We present some new ideas on how to design analogue models of quantum fields living in curved spacetimes using ultra-cold atoms in optical lattices.
  We discuss various types of static and dynamical curved spacetimes achievable by simple manipulations of the optical setup.
  Examples presented here contain two-dimensional spaces of positive and negative curvature as well as homogeneous cosmological models and metric waves. Most of them are extendable to three spatial dimensions.
  We mention some interesting phenomena of quantum field theory in curved spacetimes which might be simulated in such optical lattices loaded with bosonic or fermionic ultra-cold atoms.
  We also argue that methods of differential geometry can be used, as an alternative mathematical approach, for dealing with realistic inhomogeneous optical lattices.
\end{abstract}

\maketitle

\begin{center}
 \textit{\color{mygray}Essay written for the Gravity Research Foundation 2014 Awards for Essays on Gravitation}
\end{center}

\section{Introduction (The analogy)}

General relativists must live with the fact that curved spacetimes are not accessible in the labs -- at least the true ones. It looks different, however, when one allows for \textit{analogue} curved spacetimes.
In this short essay, we shall present a new idea on how to design analogue models of quantum fields living in curved spacetimes using ultra-cold atoms in optical lattices -- nowadays present in many quantum optical labs around the world.

As a primary goal we consider the wave equation in a curved spacetime
$ \Box_g \phi \equiv g^{\mu\nu} \nabla_\mu \nabla_\nu \phi = 0, $
where $g_{\mu\nu}$ is the metric and $\nabla_\mu$ a covariant derivative associated with it. The Greek indices $\mu, \nu$ run from $0$ to $D$, the dimension of space, where the zeroth coordinate denotes time.
Construction of the analogy suggests the choice of synchronous (or Gauss) coordinates in which the metrics satisfies the conditions $g_{00}=1$ and $g_{0i}=0$ with $i=1,...,D$.
The wave equation can be then written as
\begin{equation} \label{H-scalar_rel}
  -\d_t^2 \phi = -\Lap_g \phi = -\frac1{\sqrt{g}} \d_i \left( \sqrt{g} g^{ij} \d_j \phi \right) =: H^2 \phi,
\qquad g = \det g_{ij},
\end{equation}
where $H$, after quantization, shall become an operator generating evolution of $\phi$ in a Hilbert space.
Since the construction for relativistic fields $\phi$ is technically more involved let us, for better explanation of the main ideas, first study the non-relativistic case in which $\phi$ satisfies
\begin{equation} \label{H-scalar}
  i\d_t \phi = -\Lap_g \phi = -\frac1{\sqrt{g}} \d_i \left( \sqrt{g} g^{ij} \d_j \phi \right) =: H \phi.
\end{equation}
The analogy between ultra-cold atoms in an optical lattice and a quantum field in a curved spacetime will be established on the basis of discrete Hamiltonians governing the evolution in both systems, according to the scheme
\begin{center}
 \begin{tabular}{ccc}
   system of ultra-cold atoms & $\dashleftarrow\dashrightarrow$ & quantum field \\
   in presence of an optical potential & & in curved spacetime \\
   $\updownarrow$ & & $\updownarrow$ \\
   {\bf discrete Hamiltonian} describing &  & {\bf discrete Hamiltonian} describing \\
   atoms in an optical lattice & \hspace{1cm}$\Longleftrightarrow $\hspace{1cm} & quantum field on a lattice
 \end{tabular}
\end{center}

\subsection{The hopping Hamiltonian in an optical lattice}

\begin{figure}[!h]
  \begin{center}
    \includegraphics[height=4cm,bb=0 0 331 227]{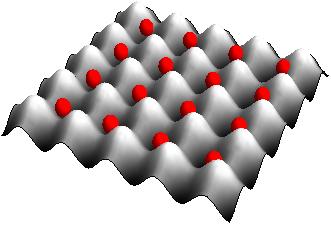}
    \hspace{0.5cm}
    \includegraphics[height=4cm,bb=0 0 538 402]{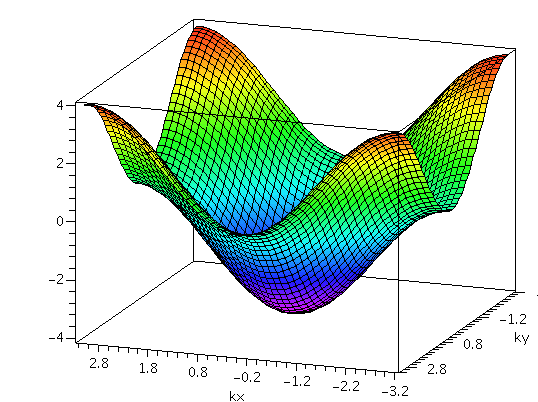}
  \end{center}
  \caption{Left: Optical potential for a 2D square lattice.
    Right: Dispersion relation for field excitations $E(\v k)$.
  \label{fig:quadratic}}
\end{figure}

As is well known in quantum optics, ultra-cold atoms moving in the presence of a periodic optical potential $ U(\v r) = V_0 \sum_{i=1}^D \sin^2(k r^i) $ created by standing laser waves with frequency $k$ (cf. Fig. \ref{fig:quadratic})
can be described by a discrete Hamiltonian upon introduction of a suitable discrete basis $\chi_n$ in the Hilbert space.
Theoretically, the most convenient choice is Wannier functions which have several useful properties: they are orthonormal, real, exponentially localized and symmetric around the minima of the periodic potential $U(\v r)$.
They provide the natural interpretation of an atom \textit{sitting} at a given minimum $\#n$, called \textit{site}, when its wavefunction is identical to the corresponding Wannier function $\chi_n$.
Since Wannier functions are not eigenfunctions of the Hamiltonian $H$, the action of the latter generates a new state which can again be decomposed in the Wannier-basis: $ H \chi_n = \sum_m T_{mn} \chi_m$. The coefficients $T_{nm}=\< \chi_n | H \chi_m\>$ are called \textit{hopping parameters} and describe the probability (more exactly: the quantum amplitude) of the tunneling of an atom from site $\#n$ to site $\#m$ in a unit of time.
The term with $n=m$ in the sum is referred to as \textit{on-site energy} are denoted $V_n := T_{nn}$. From the remaining hopping terms only those connecting the nearest neighbors $|n-m|=1$ are usually considered, while other ($|n-m|>1$) turn out to be exponentially small with growing distance.

For the second quantized field $\h \psi$ describing field excitations in the many-body language, the typical \textit{hopping Hamiltonian}, in the discussed \textit{nearest neighbor approximation}, has the form \cite{Bloch-optlat_RevNature,HubbardInOptLat}
\begin{equation} \label{H-hopping}
  \h H = \sum_{\< \v n, \v m \>} T_{\v n, \v m}\, \h \psi\s_{\v n}\, \h \psi_{\v m}
       + \sum_{\v n} V_{\v n}\, \h \psi\s_{\v n}\, \h \psi_{\v n}
\end{equation}
where we introduced a vectorial lattice-site multi-index $\v n$ which is useful when discussing regular lattices in any dimension. The first sum runs only on neighbors $\< \v n, \v m \>$ such that $|\v n - \v m| =1$.

The dispersion relation for excitations can be found by going to the momentum space via a discrete Fourier transform (from the lattice $\Z^D$ to the compact set $\B = \left[-k,k\right]^D$, called \textit{Brillouin zone}).
In the simplest case of constant hopping parameters $T_{\v n,\v m} = -J$ and $V_{\v n}=V_0$ it is
$ E(\v p)=-2J \sum_{l=1}^D \cos(p_l \la) + V_0 $
where $\v p\in\B$ is a quasi-momentum on the lattice and $\la=2\pi/k$.
In consequence, in the absence of atom--atom interactions, the ultra-cold bosonic atoms will all condense in the ground state $\v p=0$, $E=-2JD+V_0$ which does not depend on the filling (i.e. the average number of atoms per site).
The ground state excitations with small energies and quasi-momenta $|\v p|\ll k$ will behave like non-relativistic particles with
$E(\v p) \approx \frac{\v p^{\;2}}{2m} + const.$
with the effective mass $m=1/(2J\la^2)$ (cf. Fig. \ref{fig:quadratic}).

\subsection{Relation to discretized Laplacian}

The analogy becomes apparent by taking the Hamiltonian for a scalar field as in \eqref{H-scalar} and considering the underlying space as being discrete.
Discretization of $H = -\Lap$ on a regular (square or cubic) lattice with spacing $\ell$ gives
$ H \chi_{\v n} = \sum_{\v m} T_{\v n,\v m} \chi_{\v m}$ with $T_{\v n,\v n} \equiv V_{\v n} = -2D/\ell^2$ and $T_{\v n,\v m} = 1/\ell^2$ when $\<\v n, \v m\>$ are neighbors, i.e. $|\v n - \v m|=1$.
After second quantization we get
the same Hamiltonian \eqref{H-hopping} which describes hopping of ultra-cold atoms
with $J=1/\ell^2$ and $V_0=-2D/\ell^2$!
Therefore, we can say that \b{the system \eqref{H-hopping} of ultra-cold atoms in an optical lattice constitutes an analogue model for the Schrödinger field \eqref{H-scalar} living in a discrete space}. The discrete space can be treated as an approximation of the continuous space. It will, however, first become really exciting in the case of curved spacetimes which we address as next.

\section{Curved geometry in an optical lattice}


Discretization of the Laplacian $\Lap_g$, Eq. \eqref{H-scalar}, in a curved space requires more carefulness but it is a known, albeit not completely unique, procedure \cite{Oriti-Discrete_Laplacians}.
This allows for some freedom which can, here, be used for establishing the analogy with hopping Hamiltonians in the most convenient form.

We discretize the coordinates $r^i \ra r^i_{\v n}$ with the lattice multi-index $\v n\in\Z^D$ and replace the derivatives $\d_i$ by discrete differentials $\dd_i$ (similarly, as it is done in numerical computations).
In order to properly discretize the scalar field, which should later on mimic the wavefunction, we need to be conform with the conservation of the total probability in curvilinear coordinates
$
  1 = \int \bar\phi\, \phi\, \sqrt{g}\, d^2x \ra \sum_{\v n} \bar\phi_{\v n}\, \phi_{\v n}\, \sqrt{g}_{\v n},
$
where $\phi_{\v n} := \phi(r_{\v n})$ is the value of the field $\phi$ taken at the lattice point $r_{\v n}$.
On the other hand, the quantum optical system, described by the wavefunction $\psi$, does not know anything about either the curvilinear coordinates or the metric and satisfies the standard normalization condition
$
  \sum_{\v n} \bar\psi_{\v n}\, \psi_{\v n} = 1.
$
It follows the identification $\psi_{\v n} = g^{1/4}_{\v n} \phi_{\v n}$, motivated by the interpretation of the absolute value of the wavefunction squared as a measure of probability per volume element which is encoded in $\sqrt{g}$.
There is one little obstacle:
when looking at the lattice as a graph (or simplex) consisting of vortices, edges and faces,
the values of the volumes $\sqrt{g}_{\v n}$ need to be specified at the lattice points $\v n$ (graph vortices) while they are naturally defined on the lattice links (graph edges) by the (geodesic) distances between the neighboring lattice points.
This point is a bit technical but to put it briefly, $\sqrt{g}_{\v n}$ can be defined by applying some averaging procedure over neighboring links which, in the limit of small lattice distances, converges to the continuous version \cite{Oriti-Discrete_Laplacians}.

Since the discrete version of the Laplacian $\Lap_g$, Eq. \eqref{H-scalar},
would involve derivatives of the inverse metric $g^{ij}$ and its determinant $g$ and thus become too complicated for a simple analogy with the hopping Hamiltonian,
a prior integration by parts makes life much easier, according to the identity,
$
  - \int \bar\phi\, \Lap_g \phi \sqrt{g}\, d^2x = \int g^{ij} \d_i \bar\phi\, \d_j \phi \sqrt{g}\, d^2x = H.
$
Taking as a starting point the Hamiltonian, instead of the Laplacian, gives an equivalent approach (in both, continuous and discrete, cases) and leads, after discretization and second quantization, to a Hamiltonian having the hopping form \eqref{H-hopping}.
The hopping parameters $T_{\v n,\v m}$ and $V_{\v n}$ are now uniquely defined by the inverse metric $g^{ij}$
\begin{align*}
 T_{\v n, \v n+\v d} &= g^{dd}_{\v n + \v d/2}, &
 V_{\v n} &= \sum_{|\v d|=1} g^{dd}_{\v n+\v d/2}.
\end{align*}
In the hopping term $T_{\v n, \v n+\v d}$ the index $\v d$ refers to the direction and is also encoded in the corresponding component of the metric, e.g. hopping in the $x$-direction is related to $g^{xx}$.
In the on-site term $V_{\v n}$ the sum runs over all neighbors of $\v n$ and all components $g^{dd}$ of the inverse metric taken at links between $\v n$ and $\v n+\v d$, which we symbolically denote by $\v n+\v d/2$.
Since the volume elements $\sqrt{g}_{\v n+\v d/2}$ are not naturally defined on the lattice links, we can specify them by taking some mean of the values at two neighboring vortices connected by this link.
Choosing the geometric mean
$
  \sqrt{g}_{\v n+\v d/2} \ra \sqrt[4]{g}_{\v n} \sqrt[4]{g}_{\v n+\v d}
$
simplifies the further formulas essentially and
allows one to combine the volume elements $\sqrt[4]{g}$ with the scalar field $\phi$ to obtain $\psi$ as prescribed in the mapping above.
\begin{figure}[ht]
  \includegraphics[height=4cm,bb=0 0 148 157]{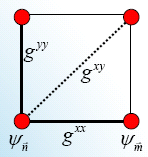}
  \vspace{-1em}
  \caption{wavefunction $\psi$ and (inverse) metric $g^{ij}$ discretized on the lattice. \label{fig:g_on_lattice}}
\end{figure}
If the metric has non-diagonal entries $g^{ij}$ with $i\neq j$ it generates corresponding hopping terms, skew on the square lattice, which are usually ignored in the nearest-neighbor approximation.
In 2D-geometry, which at the moment is the most in focus of experimental realizations, one can always find coordinates which bring any metric to the diagonal form with $g^{12}=0$ everywhere.

Summarizing, we have found a one-to-one correspondence between the hopping terms $T_{\v n,\v m}$ and the inverse metric $g^{ij}$ in the discretized space. Extrapolation to the continuous space completes the construction of \b{the analogue model of a quantum field living in curved spacetime from cold atoms hopping in an optical lattice with variable hopping terms}.
Higher probability of hopping, i.e. larger values of $T_{\v n,\v m}$, correspond to smaller distances between the adjacent lattice points in the analogue curved space, encoded in the inverse metric $g^{ij}$, and vice versa.
This opens up the possibility of creating almost any curved metric spaces by manipulation of the hopping terms in optical lattices loaded with ultra-cold atoms. Below, we will address the most interesting GR-scenarios and their experimental context.

\section{General--relativistic phenomena in optical lattices}

For particular experimental applications we will restrict these considerations to the spatial dimension $D=2$. One-dimensional lattices do not generate interesting geometries (there is no curvature in 1D), while three-dimensional lattices are more difficult to handle experimentally and although the analogy to 1+3 gravity would be more appealing, we shall give it up here for the sake of simplicity of the discussion.

\subsection{Trapping potential}

\begin{figure}[ht]
  \begin{center}
    \includegraphics[height=3cm,bb=0 0 220 220]{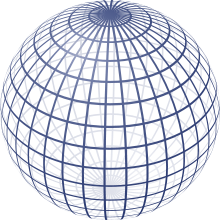}
    \hspace{1cm}
    \includegraphics[height=3cm,bb=0 0 600 412]{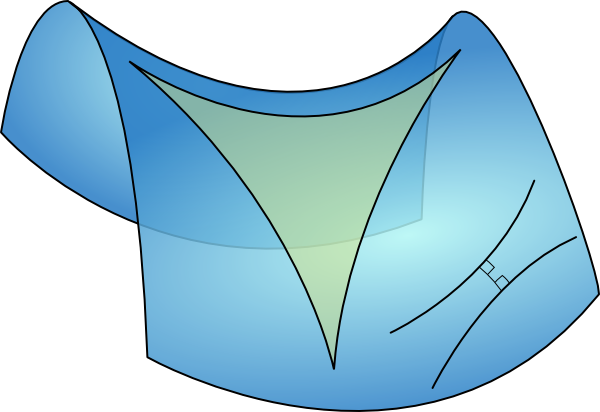}
    \hspace{1cm}
    \includegraphics[height=3cm,bb=0 0 796 462]{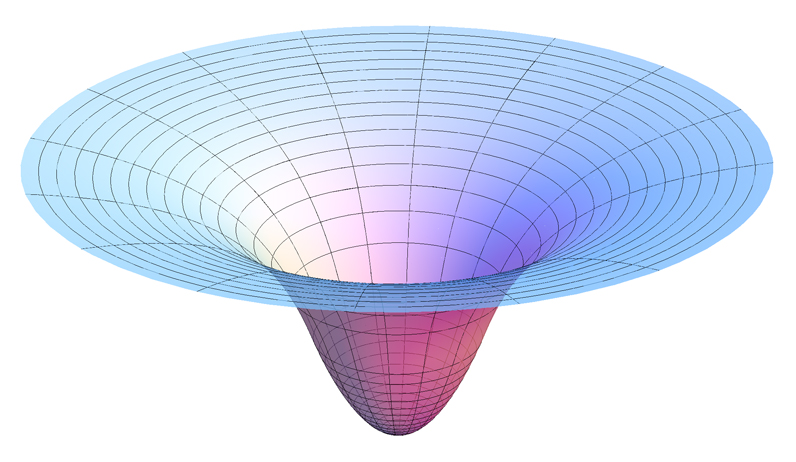}
  \end{center}
  \caption{Two surfaces of constant curvature, sphere ($K>0$, left) and hyperbolic space ($K<0$, middle), and with centrally positive curvature, asymptotically flat (right). \label{fig:2D_spaces_constK}}
\end{figure}

Although theoretical physicists like to model optical lattices as periodic and hence infinite structures the experimental reality enforces finite size of the lattice. In order to keep the atoms in finite range, additional trapping potential must be superimposed. Usually, this is a harmonic trap $U(\v r) \sim \v r^{\;2}$ but more sophisticated traps are also possible \cite{OptLat+AnharmTrap}
which have, in general, expansions
$$ U(\v r) = A \v r^{\;2} + B \v r^{\;4} + C \v r^{\;6} + ... $$
An anharmonic trap, due to its changing tilt, modifies the local hopping terms to $\tilde T_{nm} = J + \< \chi_n | U(x,y) | \chi_m \>$ where $\chi_n$ is a basis of original Wannier functions defined without the trapping potential.
Skipping the details of the calculations (which in the Wannier basis can be done analytically), it is easy to show that trap forms
$U(\v r)$ can be chosen such that $T_{nm} \approx 1 + a \v r^2 + b \v r^4 + ...$, independent of the direction of hopping.
Such a choice leads to a class of analogue curved spaces emerging from the hopping Hamiltonian with the line elements
$$ ds^2 = \frac{1}{1+ar^2+br^4} (dx^2 + dy^2). $$
These are curved surfaces with the Gaussian curvature (half of the scalar curvature)
$$ K(r) = \frac{2(a+br^2+abr^4)}{1+ar^2+br^4}. $$
Interesting cases, in which $K(r)$ becomes $r$-independent, are obtained by taking $a=2\sqrt{b}$ which leads to a constant $K>0$ representing sphere, or by taking $a=-2\sqrt{b}$ which leads to a constant $K<0$ associated with a hyperbolic (Lobachevsky) space (cf. Fig \ref{fig:2D_spaces_constK}).
Other special cases include $K(r)=2a/(1+ar^2)$ for $b=0$ (asymptotically flat) and $K(r)=2br^2/(1+br^4)$ for $a=0$ (compact).

It is striking that atoms hopping in a regularly spaced optical lattice with a trapping potential can behave as if they were moving in a curved space! 

\subsection{Variable lattice depth}

\begin{figure}[ht]
  \begin{center}
    \includegraphics[height=5cm]{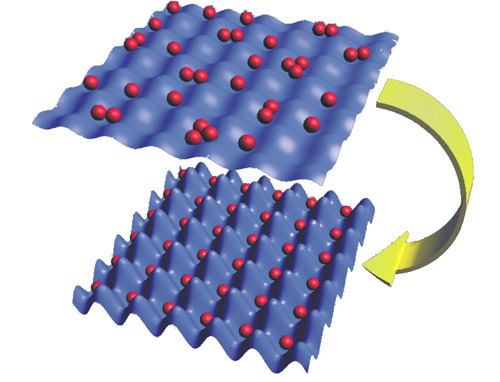}
    \hspace{1cm}
    \includegraphics[height=5cm]{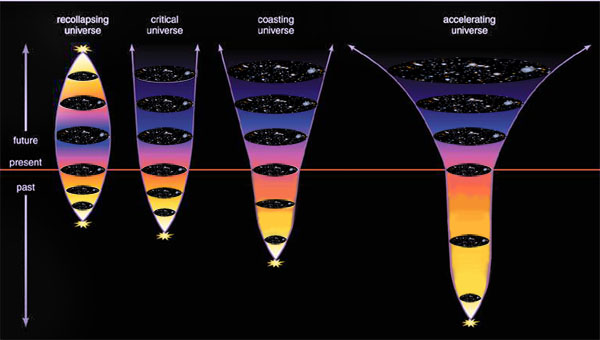}
  \end{center}
  \caption{Left: Optical lattice with changing lattice depth $A(t)$.
           Right: Cosmological expansion models (closed, flat, open, accelerated). \label{fig:optlat_A(t)}}
\end{figure}

Optical potentials can be also time-dependent. The simplest and experimentally most commonly used possibility is changing the laser intensity $A^2(t)$ of beams generating the lattice potential (cf. Fig. \ref{fig:optlat_A(t)})
$$ U(x,y,t)=A^2(t)[\sin^2(kx)+\sin^2(ky)]. $$
Varying $A(t)$ leads to time-dependent tunneling $T_{nm}(t)\sim\exp(-2A(t))$ (this analytic dependence holds for large $A$, as long as $T<2/\pi^2$)
and, consequently, to the effective line element of the emergent curved spacetime
$$ ds^2 = a^2(t) (dx^2 + dy^2), \qquad a(t) \sim \exp(2 A(t)). $$
This metric can be used for simulations of the cosmological expansion of the Universe in the flat case, $K=0$, when the cosmological scale factor $a(t)$, controlled by the lattice depth $A(t)$, is chosen properly, according to the Friedmann-Lemaître-Robertson–Walker model \cite{MTW_Gravitation}.
Other FLWR cosmological solutions, with constant positive or negative curvature $K$, can be obtained by the combination of time-dependent lattice depth $A(t)$ with the (also time-dependent) trapping potential $U(\v r)$ generating non-zero curvature $K$, as discussed above.

\subsection{Metric waves}

\begin{figure}[ht]
  \includegraphics[height=4cm,bb=0 0 662 654]{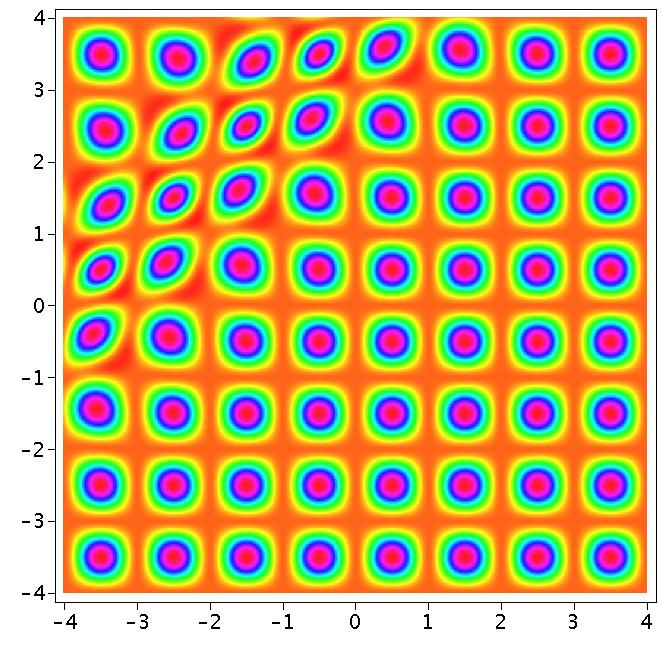}
  \hspace{0.5cm}
  \includegraphics[height=4cm,bb=0 0 662 654]{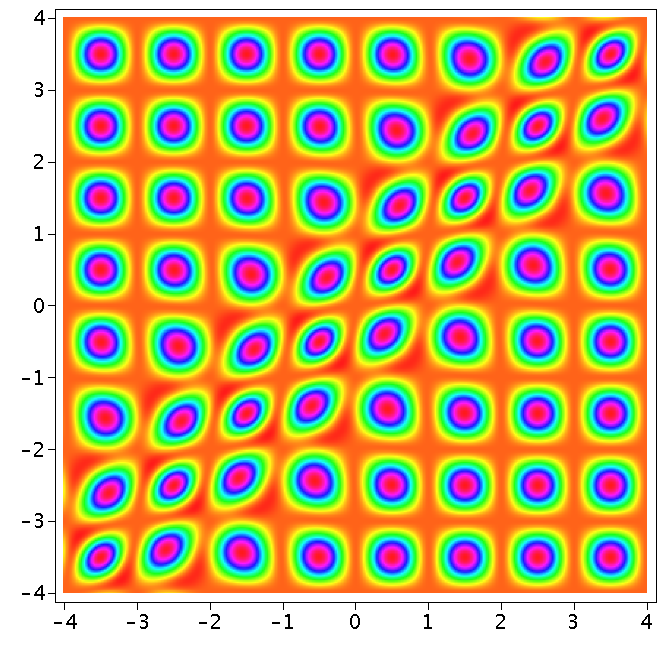}
  \hspace{0.5cm}
  \includegraphics[height=4cm,bb=0 0 662 654]{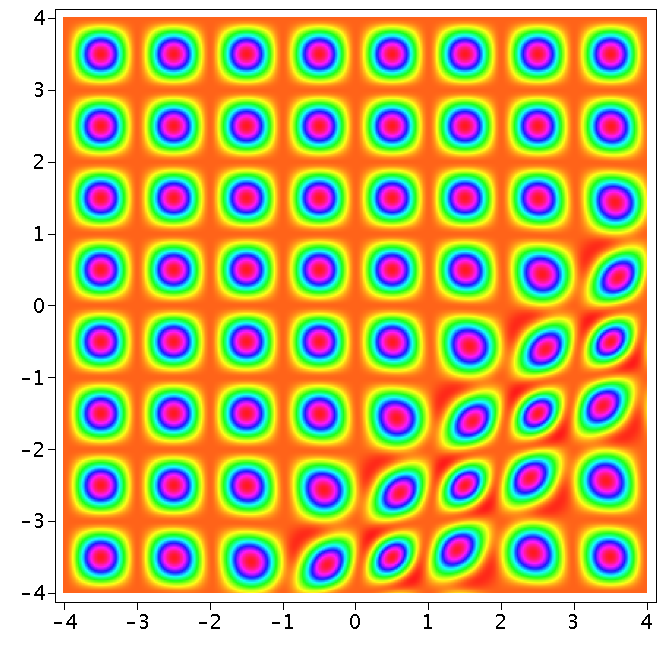}
  \caption{Left-to-right: The regular optical potential becomes deformed by a passing metric wave (from left-top to right-bottom) which carries the $\times$-polarization.
  \label{fig:optlat+GW}}
\end{figure}

Since the optical potentials can be time-dependent, it is also possible to create traveling metric waves, mimicking gravitational waves in GR. To achieve this goal additional moving laser beams are required which can locally deform the regular lattice and thus modify the hopping parameters. An example of such deformation is shown in Fig. \ref{fig:optlat+GW} where the skew hopping parameters $T^\nearrow_{\v n,\v m}(t,x,y)=-J+h(t-(x-y))$ and $T^\searrow_{\v n,\v m}(t,x,y)=-J-h(t-(x-y))$ are giving rise to a traveling $\times$-polarized metric wave \cite{MTW_Gravitation}.

\subsection{Realistic laser beams}

There is a second important aspect of the ``optical lattice -- curved space'' analogy. Instead of simulating known GR scenarios in optical lattices, existing optical lattices can be mapped onto their curved-space analogues to achieve an alternative mathematical description. In those cases where the continuous approximation of the discrete space can be applied a huge machinery of the differential geometry becomes immediately available for problems which in their  original formulations are very hard to deal with.

\begin{figure}[ht]
  \includegraphics[height=6cm]{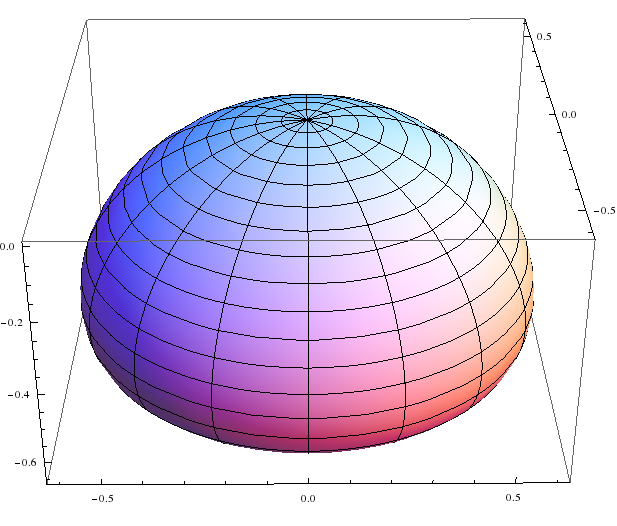}
  \hspace{1cm}
  \includegraphics[height=6cm]{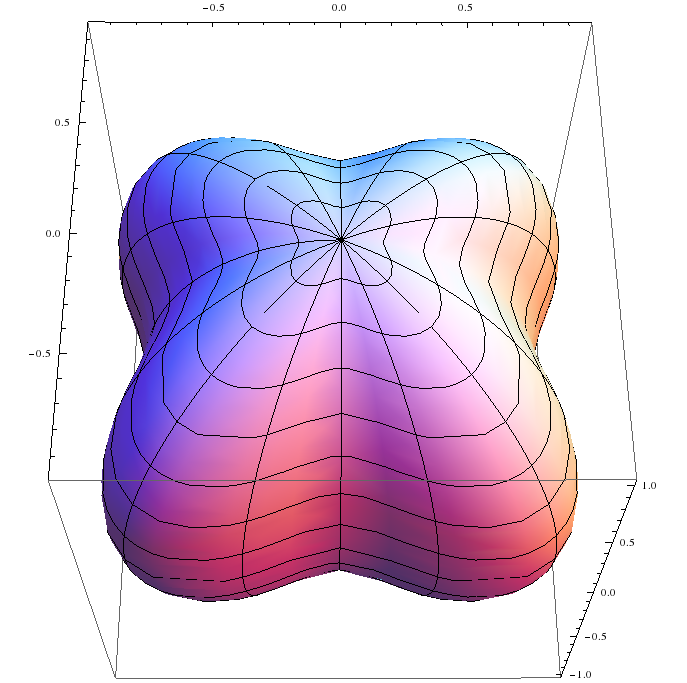}
  \caption{Effective curved spaces emerging from 2D realistic optical lattices created by laser beams of finite width. Finite 2D lattices with depth vanishing smoothly at the boundary are mapped, depending on the details, onto compact surfaces with positive curvature (left) or curvature interpolating between positive and negative (right).
  \label{fig:metrci-laser_beams}}
\end{figure}

Beyond the discussed examples of trapping potentials giving rise to curvature, there is another very natural feature of optical lattices leading to the effective curvature: laser beams have finite width and the intensity distribution inside the beam has a universal Gaussian shape \cite{Bloch-optlat_RevNature}. Therefore, the depth of almost every optical potential is spatially modulated and vanishes smoothly at the edge of the lattice, giving rise to anisotropic hopping terms:
$\tilde T^\leftrightarrow_{\v n, \v m}(\v r) \sim e^{-2 \sqrt{f(y)}}, \tilde T^\updownarrow_{\v n, \v m}(\v r) \sim e^{-2 \sqrt{f(x)}}$
with $f(s) \sim \exp(-\alpha s^2)$.
Although the hopping model is no longer adequate at such lattice edges, the system is usually combined with a trapping potential preventing the atoms from coming to the outer region. For convenience of the construction of our analogue model we can, however, include this region in the discrete space.
The outer region contains no barriers for tunneling and can be physically modeled by one quantum state extending up to infinity which the atoms can reach in finite time. Therefore, in the analog curved space picture, we expect a compact space in which the geodesic distance to coordinate-infinity is finite.
Numerical calculations confirm this picture (cf. Fig. \ref{fig:metrci-laser_beams}).

\section{Relativistic quantum fields in the lattice}

\begin{figure}[ht]
  \includegraphics[height=6cm,bb=0 0 868 782]{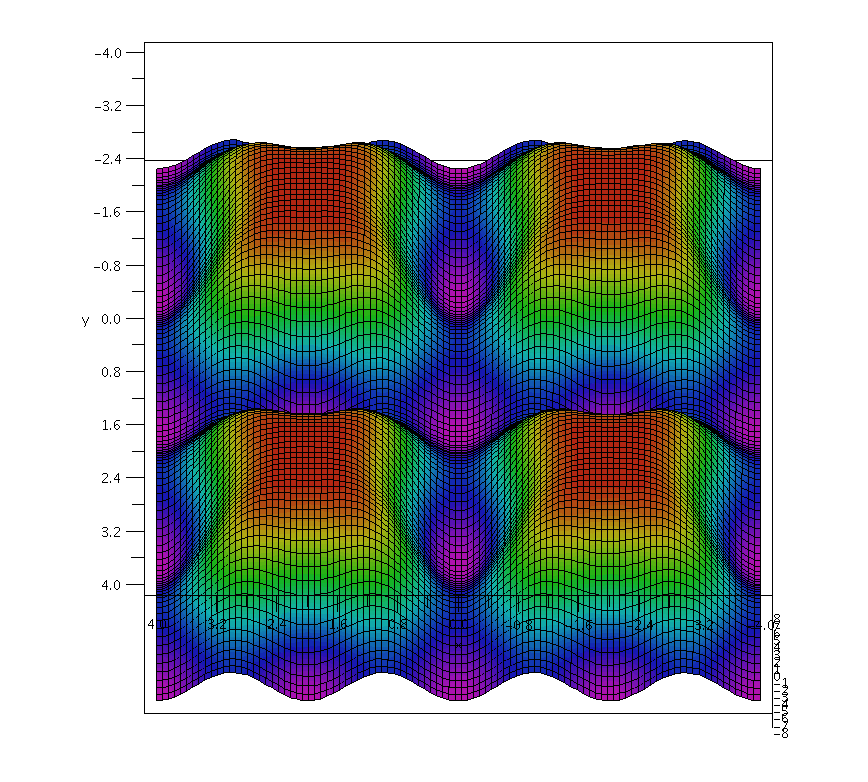}
  \hspace{1cm}
  \includegraphics[height=6cm,bb=0 0 494 520]{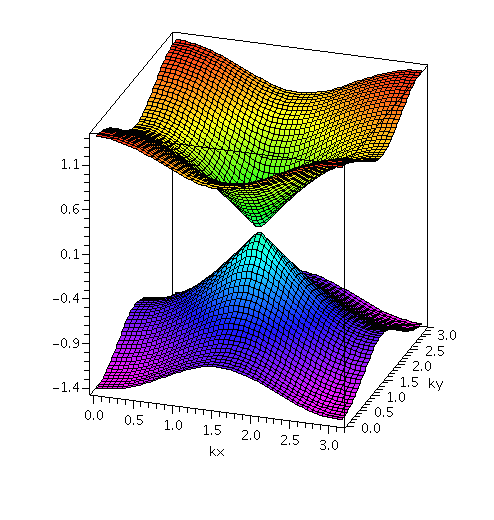}
  \caption{Left: Bichromatic 2D optical potential with three minima per \textit{supercell}. Right: Dispersion relation $E(\v p)$ for excitations of ultra-cold atoms moving in this potential landscape.
  \label{fig:bichromatic}}
\end{figure}

Finally, we must resolve the problem of relativistic quantum fields, postponed at the beginning.
Using more sophisticated forms of the optical potential, lattices with \textit{supercells} can be created, i.e. with cells having several local minima of comparable depth and thus several quantum states per elementary periodic cell within a small energy range. Such systems are effectively described by multi-component wavefunctions and can be associated with pseudo-relativistic fields. 
For instance, a bichromatic 2D optical potential, generated by superposition of laser beams with different frequencies, as depicted in Fig. \ref{fig:bichromatic}, leads to a pseudo-relativistic dispersion relation
$E(\v p)\approx \pm \sqrt{\v p^2 c^2 + m^2 c^4}$ (constants $m, c$ are effective) for excitations of a quantum field living in this potential landscape.
%
%
The corresponding evolution equation for the (discretized) field should then assume the form \eqref{H-scalar_rel}, however, here not all technical details have been clarified yet.

\section{Discussion and outlook}
It is fascinating to study the effects of quantum fields evolving in curved spacetimes -- from the classic Hawking radiation due to the horizons of black-holes to recently discovered primodal density perturbations due to inflationary cosmic expansion.
Of course, analogue models do not offer answers to the questions what the fundamental physics really is.
But they do offer answers on its behavior governed by known laws whose solutions are still not sufficiently explored.
Many models of quantum physics are known already for decades, especially those containing strongly interacting regimes, but they still cry for solutions and better qualitative understanding.
This is the domain where analogue models are expected to facilitate breakthroughs.

The models presented here become first really interesting when the simulated quantum fields are relativistic and the gravitational field (curvature) strong and/or time-dependent. Then, the true strength of quantum field theory becomes apparent, when pairs of particles and anti-particles can be dynamically created and annihilated.
To achieve this goal there is still some work ahead of us, both, on the theoretical and experimental sides.


In order to model interacting quantum fields, the interactions between ultra-cold atom must be included in the models as well as their many-body statistics (bosons or fermions).
For simulations of relativistic fermions, the Dirac equation in a curved space must be derived from the hopping Hamiltonian, what is technically more involved due to presence of spin connection terms mixing the wavefunction components.
The spinorial character of the wavefunction can be, however, easily obtained in optical lattices with honeycomb geometry in exactly the same way as it happens in graphene where electronic excitations behave effectively like relativistic fermions.
A lot of research is currently devoted to this topic and we will for sure hear about many spectacular achievements in the near future, possibly with also some impact on better understanding of curved spacetimes or even General Relativity.

\newpage

\bibliography{graphene}

\end{document}